\documentclass{raa}           

\usepackage{graphicx,times}
\usepackage{natbib}
\usepackage{amssymb,amsmath}
\bibpunct{(}{)}{;}{a}{}{,}

\usepackage[pagebackref=true]{hyperref}

\begin{document}

\title{Period-luminosity and period-luminosity-metallicity relation for $\delta$ Scuti Stars}

 \volnopage{ {\bf 20XX} Vol.\ {\bf X} No. {\bf XX}, 000--000}
   \setcounter{page}{1}

   \author{Yan-qi Liu
   \inst{1,2}, Xiao-dian Chen\inst{2,1,3,4}, Shu Wang\inst{2,1}, Kun Wang
      \inst{1},  Qi Jia\inst{1,2}, Li-cai Deng\inst{2,1,3}
   }

   \institute{ School of Physics and Astronomy, China West Normal University, Nanchong 637009, China\\
        \and
             CAS Key Laboratory of Optical Astronomy, National Astronomical Observatories, Chinese Academy of Sciences, Beijing 100101, China {\it chenxiaodian@nao.cas.cn}\\
	\and
School of Astronomy and Space Science, University of the Chinese Academy of Sciences, Beijing, 100049, China\\
\and 
Institute for Frontiers in Astronomy and Astrophysics, Beijing Normal University,  Beijing 102206, China\\
\vs \no
   {\small Received 20XX Month Day; accepted 20XX Month Day}
}

\abstract{$\delta$ Scuti ($\delta$ Sct) stars are potential distance tracers for studying the Milky Way structure. We conduct a comprehensive analysis of the period-luminosity (PL) and period-luminosity-metallicity (PLZ) relation for $\delta$ Sct stars, integrating data from the Zwicky Transient Facility (ZTF), the Transiting Exoplanet Survey Satellite (TESS), Large Sky Area Multi-Object Fiber Spectroscopic Telescope (LAMOST), Apache Point Observatory Galactic Evolution Experiment (APOGEE), and Gaia. To mitigate the impact of the Gaia parallax zero point offset, we applied a correction method, determining the optimal zero point value to be $zp_\varpi = 35 \pm 2 \, \mu\text{as}$. Using the three best bands, by varying the parallax error threshold, we found that the total error of the PLR zero point was minimized to 0.9\% at a parallax error threshold of 6\%. With this threshold, we derived the PL and PLZ relation for nine bands (from optical to mid-infrared) and five Wesenheit bands. Through our analysis, we conclude that the influence of metallicity on the PLR of $\delta$ Sct stars is not significant, and the differences across various bands are minimal.
}

   \authorrunning{Y.-Q. Liu et al. }            
   \titlerunning{PL and PLZ relation for $\delta$ Scuti Stars}  
   \maketitle

%
\section{Introduction}           
\label{sect:intro}

Delta Scuti stars ($\delta$ Sct) are short-period pulsating variables located at the intersection of the lower part of the classical Cepheid instability strip and the main sequence on the Hertzsprung-Russell diagram and they typically exhibit spectral types ranging from A2 to F5 \citep{1973A&A....23..221B,2000A&AS..144..469R,2009AIPC.1170..403H}.
The stellar mass of $\delta$ Sct stars lie in the range of approximately $1.5 \leq M \leq 2.5 M_\odot$ \citep[page~7]{2017ampm.book.....B}, with effective surface temperature between 6900 and 8900 K \citep{2011AJ....142..110M,2011A&A...534A.125U}. Their pulsation periods are generally confined to the range 0.02-0.25 days, and the typical photometric variability amplitudes in the $V$-band are less than 0.9~mag \citep{Breger_1979,1982PASP...94..845B,2000ASPC..210....3B,2000A&AS..144..469R,2013AJ....145..132C}.

The pulsations observed in $\delta$~Sct stars are primarily driven by the $\kappa$ mechanism, which arises from a resonant interaction between opacity enhancements due to helium ionization zones and the stellar temperature and density gradients \citep{1979ApJ...227..935S,2000ASPC..210....3B,1996ARA&A..34..551G}. These stars are known to exhibit complex, multimode oscillations \citep{2011A&A...534A.125U}, predominantly non-radial pressure ($p$) modes. Only a small fraction of radial modes, typically with low radial order, reach observable amplitudes \citep{1993A&AS..101..421R}, \citep[page~49]{2010aste.book.....A}, \citep[page~10]{2017ampm.book.....B}. The dominant oscillation modes that follow a well-defined period-luminosity relation (PLR) are usually the radial fundamental mode (\( n = 1, l = 0 \)) or low-order dipole modes (\( l = 1 \)) \citep{2019MNRAS.486.4348Z}.

The PLR is a fundamental astrophysical tool for distance determination. It has been extensively applied in cosmological contexts, particularly in the measurement of the Hubble constant, a key parameter governing the expansion rate of the universe \citep{2001ApJ...553...47F, 2022ApJ...934L...7R}. Beyond extragalactic distance scaling, the PLR also plays a significant role in mapping the structure of the Milky Way \citep{2019NatAs...3..320C}. While the PLR was initially developed for classical pulsators such as Cepheids and RR Lyrae stars, its applicability has progressively extended to other types of variable stars, including $\delta$~Sct stars. 
A principal challenge in establishing a precise PLR for $\delta$~Sct stars lies in their intrinsic multi-mode pulsation behavior, which complicates direct analogies with simpler radial pulsators. Early studies, such as those by \cite{1992AJ....103.1647F}, attempted to isolate stars pulsating in the fundamental mode or exhibiting dominant base frequencies to mitigate this issue. \cite{1997PASP..109..857M} further established an initial PLR for $\delta$~Sct stars, drawing parallels with the known relations for Cepheids. Subsequent work has incorporated additional complexities arising from multimode excitation and observational bandpass effects. For instance, \cite{2002ASPC..259..112L} refined the PLR for high-amplitude $\delta$~Sct stars (HADS) using parallax-based methods, including Baade–Wesselink analysis. In recent years, improvements in time-domain surveys and the advent of high-precision astrometry from Gaia have substantially advanced both pulsation mode identification and luminosity calibration, thereby enabling significant refinement of the $\delta$~Sct PLR \citep{2019MNRAS.486.4348Z, J20, 2021PASP..133h4201P, 2022MNRAS.516.2080B, 2022ApJ...940L..25M, 2023aa...674A..36G, 2023arXiv230915147S,2023AJ....165..190N, 2024ApJ...972..137G}.

Metallicity has long been recognized as a significant factor influencing the PLR. It affects both the intrinsic luminosity and pulsation properties of variable stars, and therefore must be considered when deriving accurate distances from PL-based methods. In the context of $\delta$~Sct stars, metallicity modulates not only the stellar brightness but also the pulsation period \citep{2021AcA....71..189S}. For a given absolute magnitude, higher metallicity tends to correlate with longer pulsation periods, thereby altering both the slope and the zero-point of the PLR. Consequently, failing to correct for metallicity effects may introduce systematic biases in distance estimates \citep{2004ASPC..310..525M,2007AJ....133.2752M,2011AJ....142..110M}. \citet{2020MNRAS.493.4186J} analyzed a large sample of $\delta$~Sct stars from the ASAS-SN survey to examine the period–metallicity relation, finding only a weak correlation, with a metallicity dispersion of approximately $\sigma \sim 0.25$ dex.

In this study, we aim to calibrate the PL and PLZ relation of $\delta$ Sct stars from optical to mid-infrared band, using a larger sample of $\delta$ Sct stars. The paper is organized into six sections: Section \ref{sec:2} details the origin of the sample and the selection criteria, Section \ref{sec:3} describes the methodologies used, Section \ref{sec:4} presents the derived PL and PLZ relation, as well as the corresponding fitting parameters, Section \ref{sec:5} discusses the results; and Section \ref{sec:6} concludes with a summary of our findings.
\section{DATA} \label{sec:2}

\subsection{Sample of stars}\label{sec:sample}
The Zwicky Transient Facility (ZTF) is a 48-inch Samuel Oschin Schmidt telescope located at the Palomar Observatory, equipped with a wide field of view of 47 deg$^2$. ZTF is designed to provide high-cadence time-series observations, delivering approximately 450 epochs for 1.8 billion astronomical sources.  Its primary scientific goals include the detection and characterization of transient events, studies of stellar variability, and investigations of solar system objects \citep{2019PASP..131g8001G, 2019PASP..131a8003M}. Based on the ZTF Data Release 2 (DR2), which spans a baseline of approximately 470 days, \citet{2020ApJS..249...18C} compiled a catalog containing around 780,000 periodic variable stars. From this catalog, we selected 16,709 sources classified as $\delta$~Sct stars for further analysis.

The Transiting Exoplanet Survey Satellite (TESS) is a space-based observatory designed to detect transiting exoplanets around nearby bright stars \citep{2014SPIE.9143E..20R,2015JATIS...1a4003R}. The TESS instrument comprises four wide-field CCD cameras, collectively covering a field of $24^\circ \times 96^\circ$ on the sky. Each camera monitors the brightness of approximately 15,000–20,000 stars with a 2-minute cadence, over observing durations ranging from 27 to 356 days. In addition, 30-minute full-frame images (FFIs) are recorded continuously for broader sky coverage. Owing to its continuous monitoring and large sky coverage, TESS is particularly well-suited for the study of variable stars. Recently, \citet{2025ApJS..276...57G} published a catalog of 72,505 variable sources based on TESS observations, which includes 9,012 $\delta$~Sct stars.

To construct a comprehensive dataset for our analysis, we combined the $\delta$~Sct star catalogs from ZTF and TESS. Cross-matching the two datasets resulted in 1,394 common sources. A comparison of the pulsation periods derived independently from each survey for these overlapping stars shows strong agreement, validating the consistency and reliability of both datasets. After removing duplicates, we obtained a final sample of 24,327 unique $\delta$~Sct stars, which serves as the basis for the subsequent analysis in this study.

\subsection{Metallicity data}\label{sec2.2}
We first obtained spectroscopic metallicity measurements from the Data Release 11 (DR11) of the Large Sky Area Multi-Object Fiber Spectroscopic Telescope (LAMOST; \citealt{2012RAA....12.1197C}). Cross-matching our $\delta$~Sct star sample with the LAMOST DR11 catalog yielded 1,686 stars with available [Fe/H] values. For the remaining 21,244 stars without LAMOST metallicity, we further cross-matched this subset against the Data Release 17 (DR17) of the Apache Point Observatory Galactic Evolution Experiment (APOGEE; \citealt{2011AJ....142...72E, 2017AJ....154...94M}). This supplementary cross-match provided metallicity estimates for an additional 21 stars. By combining the results from LAMOST and APOGEE, we assembled a final spectroscopic metallicity sample consisting of 1,707 $\delta$~Sct stars. The [Fe/H] values span a range from $-2.48$ to $+0.64$ dex, with a mean metallicity of approximately $-0.28$ dex. 

\subsection{Multiband data}\label{sec2.3}

We further cross-matched the 1,707 $\delta$~Sct stars with the Gaia Data DR3 \citep{2023A&A...674A...1G}) catalog to obtain precise astrometric parameters and broad-band photometry in the \textit{G}, \textit{BP}, and \textit{RP} bands. In addition, we acquired near- and mid-infrared photometry by matching these sources with the Two Micron All Sky Survey (2MASS; \citealt{2006AJ....131.1163S}) and the Wide-field Infrared Survey Explorer (WISE; \citealt{2010AJ....140.1868W}) catalogs, yielding magnitudes in the $J$, $H$, $K_S$, $W_1$, $W_2$, and $W_3$ bands.

Extinction corrections were applied using relative extinction coefficients provided by \citet{2019ApJ...877..116W}. The line-of-sight reddening values $E(B-V)$ were obtained via the \texttt{dustmaps} package \citep{2018JOSS....3..695M}, based on the `SFD' dust map \citep{2011ApJ...737..103S}. To mitigate the effects of interstellar extinction on our photometric analysis, we also computed extinction-independent Wesenheit magnitudes \citep{1982ApJ...253..575M}, which are commonly used in PLR studies due to their reduced sensitivity to reddening. These Wesenheit indices are as follows: $G-1.89(BP-RP)$, $W_1-0.094(BP-RP)$, $W_2-2.032(W_1-W_2)$, $J-0.686(K_S-J)$, $K_S-1.464(H-K_S)$ where the coefficient \( R \) is selected based on the photometric bands involved.

\subsection{Data Filtering}\label{sec2.4}

After obtaining the full set of photometric and astrometric data, we performed a rigorous quality filtering process to ensure the reliability of the final sample. We first excluded sources with Renormalized Unit Weight Error (RUWE) values greater than 1.4 in Gaia DR3, a commonly adopted threshold to remove sources with potentially poor astrometric solutions. Next, we applied a parallax signal-to-noise filter by imposing a fractional parallax uncertainty threshold, initially set to 20\%. This threshold was iteratively refined in accordance with the criteria described in Section~\ref{sec:3}, progressively tightening the sample selection. After applying these cuts and correcting Gaia parallaxes using the \texttt{zero-point} Python package \citep{lindegren2021gaiaads}, we obtained a sample of 1,071 stars with high-quality astrometric and photometric parameters.

\begin{figure*}[!ht]
\centering
\includegraphics[width=1\textwidth]{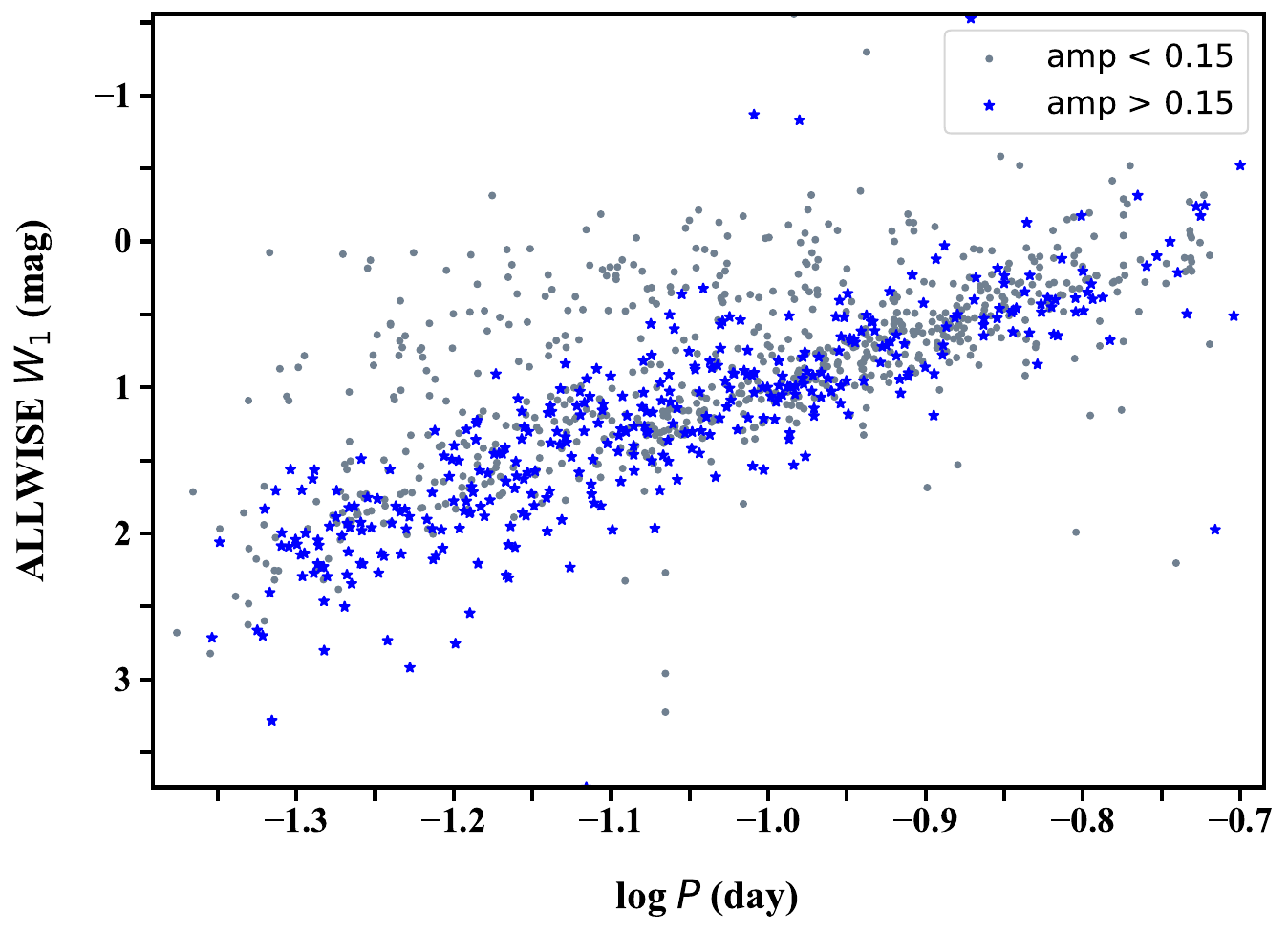}
\caption{The figure compares the distributions of 1,071 $\delta$ Sct stars on the PLR diagram, contrasting samples with amplitude limited to a minimum of 0.15 mag (represented by stars) and samples without amplitude restriction (represented by dots).}
    \label{f1}
\end{figure*}

To further isolate high-confidence fundamental-mode pulsators, we applied an amplitude-based selection criterion. As demonstrated in Figure~7 of \citet{2024ApJS..273....7J}, first-overtone (1O) $\delta$ Sct stars typically exhibit photometric amplitudes smaller than 0.15~mag. Additionally, \citet{2025arXiv250320557J} discussed that samples with amplitudes below this threshold are more likely to suffer from 1O-mode contamination based on the PLR. We examined our 1,071 candidates in the PLR diagram and found that a significant fraction of stars with amplitudes less than 0.15~mag align with the expected 1O-mode PLR (see Figure \ref{f1}). Therefore, we retained only stars with amplitudes greater than 0.15~mag. The resulting high-confidence sample consists of 354 $\delta$~Sct stars and is used for all subsequent PLR analysis.

\section{METHOD}\label{sec:3}
Although we have corrected for potential systematic biases in the Gaia parallax data, residual biases in the corrected Gaia parallaxes still persist \citep{Riess_2019, 2021ApJ...911L..20R}, which may significantly affect the accuracy of the PLR. Therefore, to further improve the precision of the PLR derivation, additional corrections for the residual offsets in Gaia parallaxes are necessary. We employ the method proposed by \cite{Riess_2019,Riess_2022} to express the absolute magnitude as
\begin{equation}\label{equ2}
    W = a_1 (\log P - \log P_0) + a_2 + a_3 \text{[Fe/H]},
\end{equation}
where \( \log P_0 \) denotes the logarithm of the mean value of the period, which is -1.031 in our sample, \( a_1 \) and \( a_2 \) represent the slope and intercept of the PLR, respectively, and \( a_3 \) is the coefficient for the metallicity term. 
Subsequently, Equ. \ref{equ3} defines an extinction-free distance modulus.
\begin{equation}\label{equ3}
    \mu_{0,i} = m_{W,i} - W
\end{equation}
Using the distance modulus, the photometric parallax is further defined as
\begin{equation}\label{equ4}
\pi_{\text{phot},i} = 10^{-0.2(\mu_{0,i} - 10)}
\end{equation}
The discrepancy between the photometric distance and the Gaia parallax, which we corrected in Section \ref{sec2.2}, represents the residual offset that needs to be determined. The specific value of this offset $zp_\varpi$ can be calculated using Equ. \ref{equ5}.

\begin{equation}\label{equ5}
\varpi = 10^{-0.2(m_W + a_1(\log P - \log P_0) + a_2 + a_3\rm [Fe/H]) + 2} + zp_\varpi, \quad  
\end{equation} 

Due to the non-uniform distribution of parallax uncertainties in Gaia data across different magnitudes, colors, and spatial positions, it is essential to mitigate the influence of high-uncertainty measurements in order to improve the overall fitting accuracy. To address this, the uncertainty defined in Equ. \ref{equ6} is used to calculate the weights in our fitting procedure:
\begin{equation}\label{equ6}
e_{\rm para}^2 = {\sigma_\varpi}^2 + \varpi^2 \times \frac{{0.16^2 + \sigma_m^2}}{4 \times 1.086^{2}}. \quad
\end{equation}
Here, \(\sigma_m\) represents the apparent magnitude error, and 0.16 denotes the assumed intrinsic scatter of the PLR. We apply Equ. \ref{equ10} to calibrate the parallax zero-point offset using these weights. In this equation, $\pi_{\mathrm{DR3},i}$ the corrected Gaia DR3 parallax of the $i^\text{th}$ star, $\pi_{\mathrm{phot},i}$ is the corresponding photometric parallax, $zp_\varpi$ is the zero-point offset, and $e_{i, \rm para}$ is the weight.
\begin{equation}\label{equ10}
\chi^2 = \sum \frac{(\pi_{\mathrm{DR3},i} - \pi_{\mathrm{phot},i} - zp_\varpi)^2}{e_{i, \rm para}^2}
\end{equation}

\section{Results}\label{sec:5}

\subsection{Gaia parallax zero-point offset}
\begin{figure*}[!ht]
\centering
\includegraphics[width=1\textwidth]{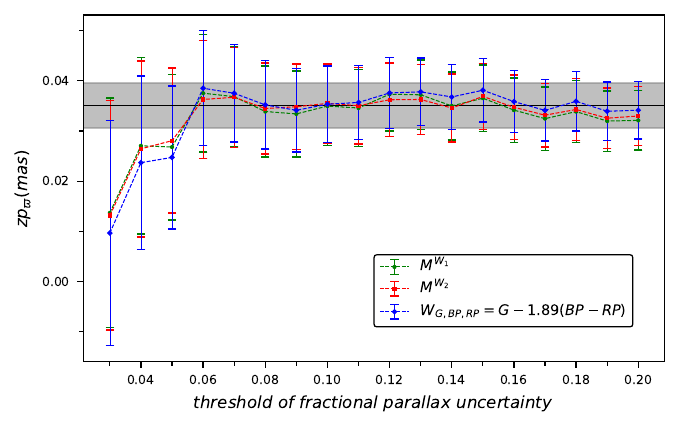}
\caption{The figure illustrates the parallax zero point values \( zp_\varpi \) for three bands (\( W_1 \), \( W_2 \) and \( W_{G,BP,RP} \)) across a range of parallax uncertainty thresholds from 2\% to 20\%. The shaded gray region represents the range of three times the standard deviation of \( zp_\varpi \), indicating the statistical uncertainty of the zero point values. The black line indicates \( zp_\varpi \) = 35 \( \mu \text{as} \), which serves as the reference zero point value for all bands.}
    \label{f2}
\end{figure*}

The choice of parallax error ratio threshold has a direct impact on the numerical stability of the zero-point offset estimation. To assess its influence, we systematically varied the fractional parallax uncertainty threshold from 2\% to 20\%, and evaluated its effect on the derived zero-point values \( zp_\varpi \) (see Figure~\ref{f2}). Our results show that within the threshold range of 6\% to 20\%, the \( zp_\varpi \) values remain relatively stable and are associated with smaller uncertainties. In contrast, when the threshold is more stringent, the estimated 
\( zp_\varpi \) values exhibit larger fluctuations and increased uncertainty, primarily due to the significantly reduced sample size in this regime. Based on this analysis, we conclude that a parallax error threshold between 6\% and 20\% offers an optimal balance between sample size and statistical stability for \( zp_\varpi \) estimation. To further refine the parallax zero-point correction, we focused on photometric bands that are minimally affected by extinction and exhibit intrinsically low dispersion in the PLR. Specifically, we used the Wesenheit magnitudes derived from the \( W_1 \), \( W_2 \), and \( W_{G,BP,RP} \) bands to determine the \( zp_\varpi \) value. The standard deviation of \( zp_\varpi \)was computed for each band, and the final adopted parallax zero-point correction was derived by averaging the results across the three bands, yielding: \( zp_\varpi = 35 \pm 2 \, \mu\text{as} \).

\subsection{Derived relations} \label{sec:4}
\begin{figure}[htbp]
\centering
\includegraphics[width=1\textwidth]{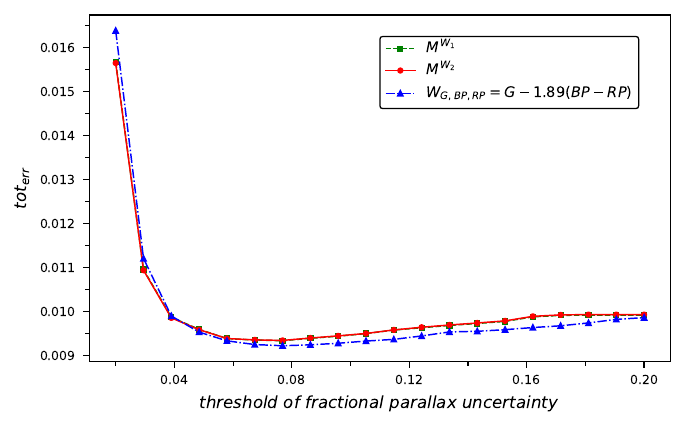}
\caption{Total PLR zero point error under different parallax uncertainties.}
\label{f3}
\end{figure}

After adopting a fixed Gaia parallax zero-point correction of $zp_\varpi = 35 \pm 2 \, \mu\text{as}$, we proceeded to determine the optimal parallax error threshold for fitting the parameters of the PLR. To achieve this, we calculated the total uncertainty in the PLR zero point under varying parallax error thresholds, with the objective of minimizing this value. The total error was computed using Equation~\ref{equ7}, where \( a_{2,std} \) denotes the standard deviation of the intercept term in the PLR fit, and the second term represents the contribution from parallax uncertainty propagation. Figure~\ref{f3} displays the behavior of the total zero-point error as a function of the parallax error threshold, for three representative photometric bands: $W_1$, $W_2$, and $W_{G,BP,RP}$, under the fixed \( zp_\varpi \). The total error initially decreases with a stricter parallax threshold, reaching a minimum and then rising again due to shrinking sample size. According to our analysis, the total error reaches its lowest value (0.9\%) at a parallax error threshold of 6\%. Therefore, we adopt 6\% as the optimal fractional parallax uncertainty threshold for subsequent PLR determination.
\begin{equation}\label{equ7}
    tot_{err} = \sqrt[2]{\left(\frac{a_{2,std}}{2/1.086}\right)^2 + \left(\frac{2}{\text{mean}(\bar{\omega})/1000}\right)^2}
\end{equation}

\begin{figure}[ht!]
\centering 
\includegraphics[width=1\textwidth]{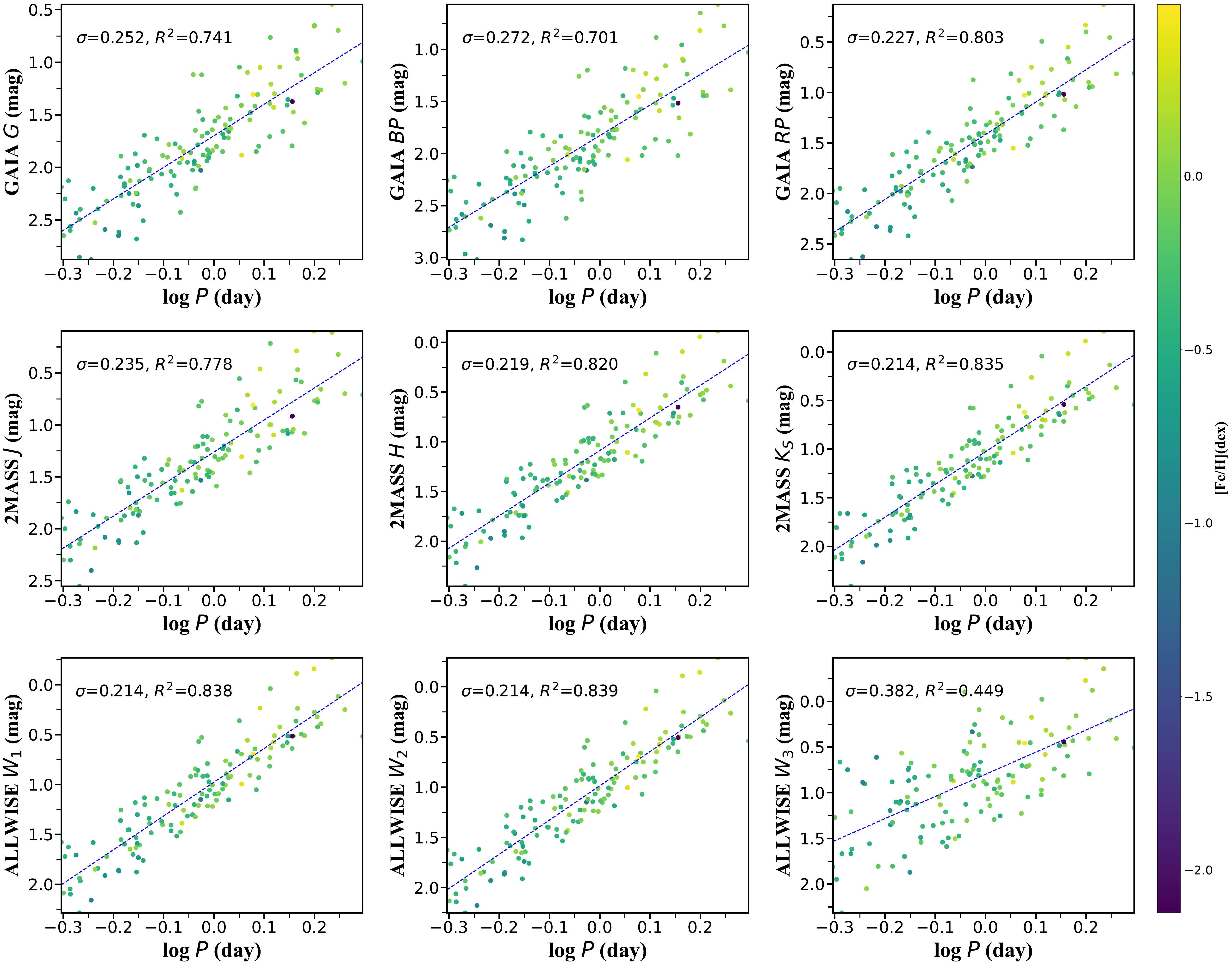} 
\caption{PLR for $\delta$ Sct stars in various photometric bands. The data points represent the observed absolute magnitudes, and the fitted lines are the results of the linear fit. The LAMOST metallicities are coded with color.} 
\label{f4} 
\end{figure}

\begin{figure}[ht!] 
\centering 
\includegraphics[width=1\textwidth]{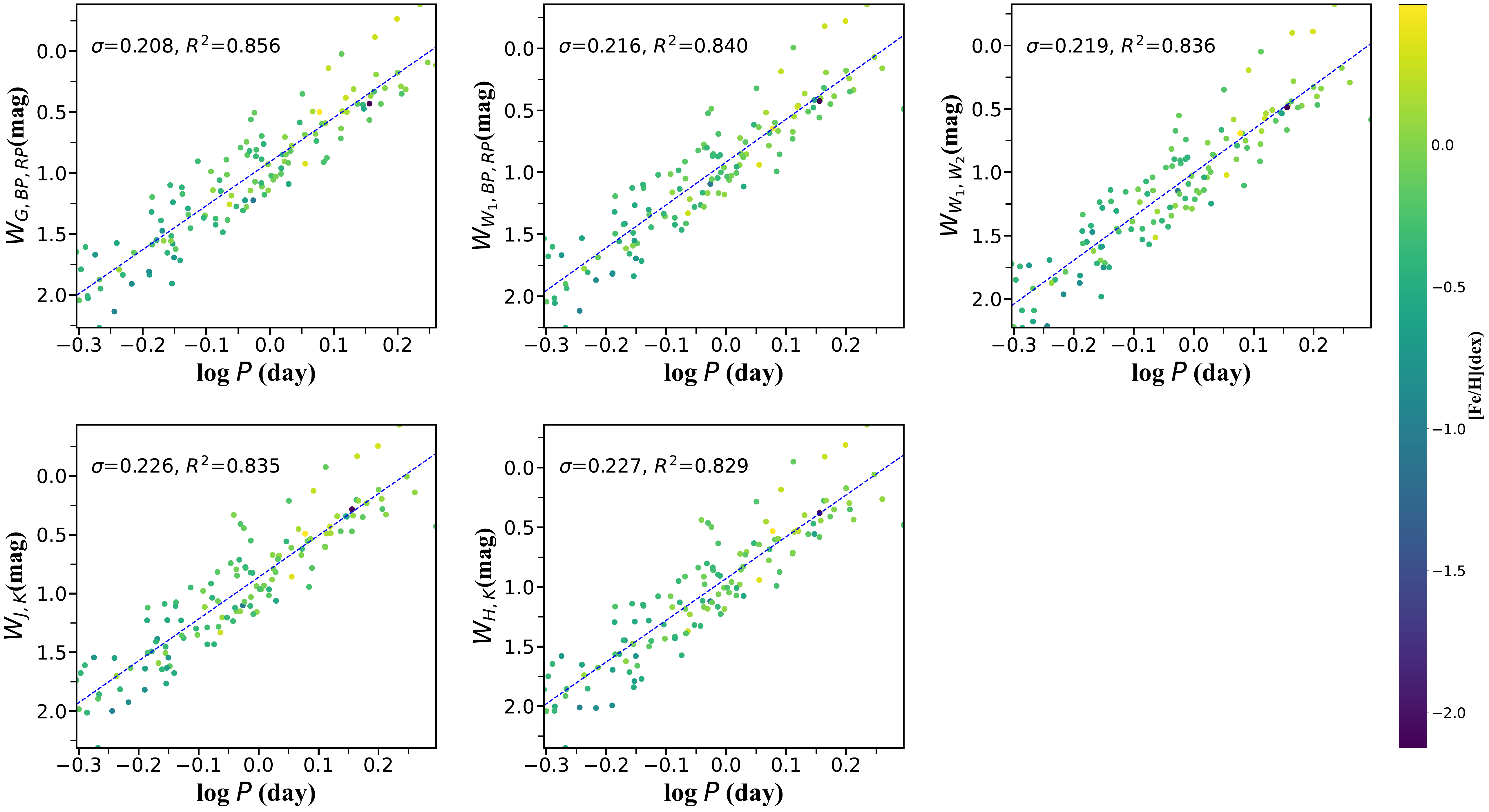} 
\caption{Similar to Figure \ref{f3}, but for the PW relation of $\delta$ Sct stars based on the Wesenheit magnitudes.} 
\label{f5} 
\end{figure}

Using this criterion, we selected the subsample of stars with parallax uncertainties below 6\% and constructed their PL, PW, PLZ, and PWZ relations using extinction-corrected absolute magnitudes and Wesenheit magnitudes as described in \ref{sec2.2}. The analysis includes nine single-band relations ($G$, $BP$, $RP$, $J$, $H$, $K_S$, $W_1$, $W_2$, $W_3$) and five Wesenheit-band relations ($W_{G,BP,RP}$, $W_{W_1,BP,RP}$, $W_{W_1,W_2}$, $W_{J,K_S}$, $W_{H,K}$), yielding a total of 14 band combinations. All parameter fittings were performed in the parallax space. First, a linear regression preprocessing was conducted, where data with residuals beyond $3\sigma$ were removed. Then, the least squares fitting model outlined in Section \ref{sec:3} was applied, considering error weighting, to obtain the parameters and their standard deviations. Figure \ref{f4} presents the PLR for the nine single-band samples, while Figure \ref{f5} illustrates the PLR for the Wesenheit magnitudes. In these figures, the color bars represent the metallicity data of the sample stars, $\sigma$ denotes the dispersion of the PLR, and $R^2$ is the coefficient of determination for the fits. Table \ref{tab1} summarizes the fitting results for the PL and PW relation across different bands, including the slope $a_1$, intercept $a_2$, metallicity term coefficient $a_3$, dispersion $\sigma$, the number of samples used for fitting $n$, and the total zero-point error.  
\begin{equation}\label{equ8}
    W = a_1 (\log P - \log P_0)  + a_2,
\end{equation}
\begin{table}
    \scriptsize
    \centering
    \caption{PL and PLZ relation for $\delta$ Sct determined based on a Gaia parallax zero point of \( zp_\varpi = 35\pm 2\) $\mu$as.}
    \begin{tabular}{ccccccccc}
    \hline 
     & Filters & $a_1$ & $a_2$ & $a_3$ & Total zp error $tot_{err}$ (\%) & $\sigma$ (mag) & $R^2$ & n\\
    \hline
    \multicolumn{8}{c}{PLR, $\log P_0 = -1.031$} \\
    & $ G $ & $-3.003 \pm 0.155$ & $1.795 \pm 0.022$ &  & $0.97$ & $0.252$ & $0.741$ &$141$ \\
    & $ BP $ & $-2.935 \pm 0.167$ & $1.922 \pm 0.024$ &  & $ 0.98 $ & $0.272$ & $0.701$ &$141$ \\
    & $ RP $ & $-3.213 \pm 0.139$ & $1.518 \pm 0.020$ &  & $ 0.97 $ & $0.227$ & $0.803$ &$141$ \\
    & $ J $ & $-3.084 \pm 0.145$ & $1.356 \pm 0.021$ &  & $ 0.98 $ & $0.235$ & $0.778$ &$138$ \\
    & $ H $ & $-3.277 \pm 0.135$ & $1.188 \pm 0.019$ &  & $ 0.98 $ & $0.219$ & $0.820$ &$139$ \\
    & $ K $ & $-3.364 \pm 0.132$ & $1.131 \pm 0.019$ &  & $ 0.98 $ & $0.214$ & $0.835$ &$139$ \\
    & $ W_1 $ & $-3.386 \pm 0.132$ & $1.081 \pm 0.019$ &  & $ 0.98 $ & $0.214$ & $0.838$ &$138$ \\
    & $ W_2 $ & $-3.408 \pm 0.132$ & $1.091 \pm 0.019$ &  & $ 0.98 $ & $0.214$ & $0.839$ &$138$ \\
    & $ W3 $ & $-2.423 \pm 0.237$ & $0.873 \pm 0.034$ &  & $ 1.45 $ & $0.382$ & $0.449$ &$105$ \\
    & $ W_{G,BP,RP} $ & $-3.615 \pm 0.130$ & $1.023 \pm 0.018$ &  & $ 0.97 $ & $0.227$ & $0.856$ &$141$ \\
    & $ W_{W_1,BP,RP} $ & $-3.457 \pm 0.133$ & $1.024 \pm 0.019$ &  & $ 0.97 $ & $0.276$ & $0.840$ &$138$ \\
    & $ W_{W_1,W_2} $ & $-3.453 \pm 0.135$ & $1.111 \pm 0.019$ &  & $ 0.97 $ & $0.208$ & $0.836$ &$138$ \\
    & $ W_{J,K_S} $ & $-3.555 \pm 0.139$ & $0.975 \pm 0.020$ &  & $ 0.98 $ & $0.216$ & $0.688$ &$138$ \\
    & $ W_{H,K_S} $ & $-3.499 \pm 0.140$ & $1.037 \pm 0.020$ &  & $ 0.98 $ & $0.219$ & $0.829$ &$138$ \\
    \multicolumn{8}{c}{PLZ relation, $\log P_0 = -1.031$} \\
    & $ G $ & $-2.801 \pm 0.162$ & $1.749 \pm 0.027$ & $ -0.229 \pm 0.071$ & $ 1.24 $ & $0.242$ & $0.760$ &$141$ \\
    & $ BP $ & $-2.730 \pm 0.175$ & $1.874 \pm 0.029$ & $ -0.233 \pm 0.077$ & $ 1.26 $ & $0.262$ & $0.720$ &$141$ \\
    & $ RP $ & $-3.020 \pm 0.150$ & $1.470 \pm 0.025$ & $-0.217 \pm 0.066$ & $ 1.22 $ & $0.225$ & $0.806$ &$141$ \\
    & $ J $ & $-3.019\pm0.134$ & $1.266\pm0.024$ & $-0.198\pm0.058$ & $ 1.22 $ & $0.229$ & $0.788$ &$138$ \\
    & $ H $ & $-3.204\pm0.134$ & $1.118\pm0.024$ & $-0.161\pm0.058$ & $ 1.22 $ & $0.216$ & $0.825$ &$139$ \\
    & $ K $ & $-3.331\pm0.134$ & $1.059\pm0.024$ & $-0.140\pm0.058$ & $ 1.21 $ & $0.218$ & $0.830$ &$139$ \\
    & $ W_1 $ & $-3.301\pm0.133$ & $1.007\pm0.024$ & $-0.174\pm0.058$ & $ 1.21 $ & $0.219$ & $0.831$ &$138$ \\
    & $ W_2 $ & $-3.332\pm0.133$ & $1.019\pm0.024$ & $-0.169\pm0.058$ & $ 1.21 $ & $0.220$ & $0.830$ &$138$ \\
    & $ W3 $ & $-3.029\pm0.220$ & $0.647\pm0.034$ & $-0.160\pm0.077$ & $ 1.64 $ & $0.394$ & $0.450$ &$105$ \\
    & $ W_{G,BP,RP} $ & $-3.408\pm0.131$ & $0.953\pm0.023$ & $-0.210\pm0.056$ & $ 1.20 $ & $0.207$ & $0.855$ &$141$ \\
    & $ W_{W_1,BP,RP} $ & $-3.372\pm0.132$ & $0.953\pm0.024$ & $-0.171\pm0.057$ & $ 1.20 $ & $0.225$ & $0.827$ &$138$ \\
    & $ W_{W_1,W_2} $ & $-3.392\pm0.132$ & $1.040\pm0.024$ & $-0.157\pm0.057$ & $ 1.20 $ & $0.230$ & $0.820$ &$138$ \\
    & $ W_{J,K_S} $ & $-3.447 \pm 0.149$ & $0.943 \pm 0.025$ & $-0.123 \pm 0.066$ & $ 1.21 $ & $0.275$ & $0.688$ &$138$ \\
    & $ W_{H,K_S} $ & $-3.414\pm0.132$ & $0.979\pm0.024$ & $-0.163\pm0.056$ & $ 1.20 $ & $0.226$ & $0.829$ & $138$ \\
\hline
    \end{tabular}
    \label{tab1}
\end{table}

According to the data analysis presented in Table~\ref{tab1}, the $\sigma$ values of the nine single bands gradually decrease from the optical to the mid-infrared bands, while the $R^2$ correlation coefficients progressively increase, indicating that the PLR becomes more stable in this process. Among them, the $W_2$ band exhibits the best performance, with an $R^2$ value of 0.839 and a $\sigma$ value of 0.214. In the Wesenheit bands, the $(W_{G,BP,RP})$ band is the most optimal, with an $R^2$ value of 0.856 and a $\sigma$ value of 0.208. Furthermore, the slope of the PLR tends to increase from the optical to the mid-infrared bands, while the intercept gradually decreases. The coefficient of the metallicity term has a relatively large error, with significance ranging from 1.86 to 3.75$\sigma$, suggesting that the PLR of $\delta$ Sct stars is weakly influenced by metallicity.

\subsection{Metallicity's influence on the PL/PW relation}
To further investigate the influence of metallicity on the PLR, we analyzed the residuals of absolute magnitudes across different metallicity intervals. The residuals are calculated by subtracting the fitted PLR values from the observed absolute magnitudes. Smaller residuals indicate a closer agreement between the model predictions and the observed values, suggesting a better fit. If the PLR is significantly affected by metallicity, a correlation between the residuals and metallicity is expected. We fitted the residuals of the PLR against metallicity for nine individual photometric bands and five Wesenheit bands, considering a metallicity range from [Fe/H] = $-1.2$ to $+0.5$ dex, which was uniformly divided into ten intervals. 

\begin{figure}[ht!] 
\centering \includegraphics[width=1\textwidth]{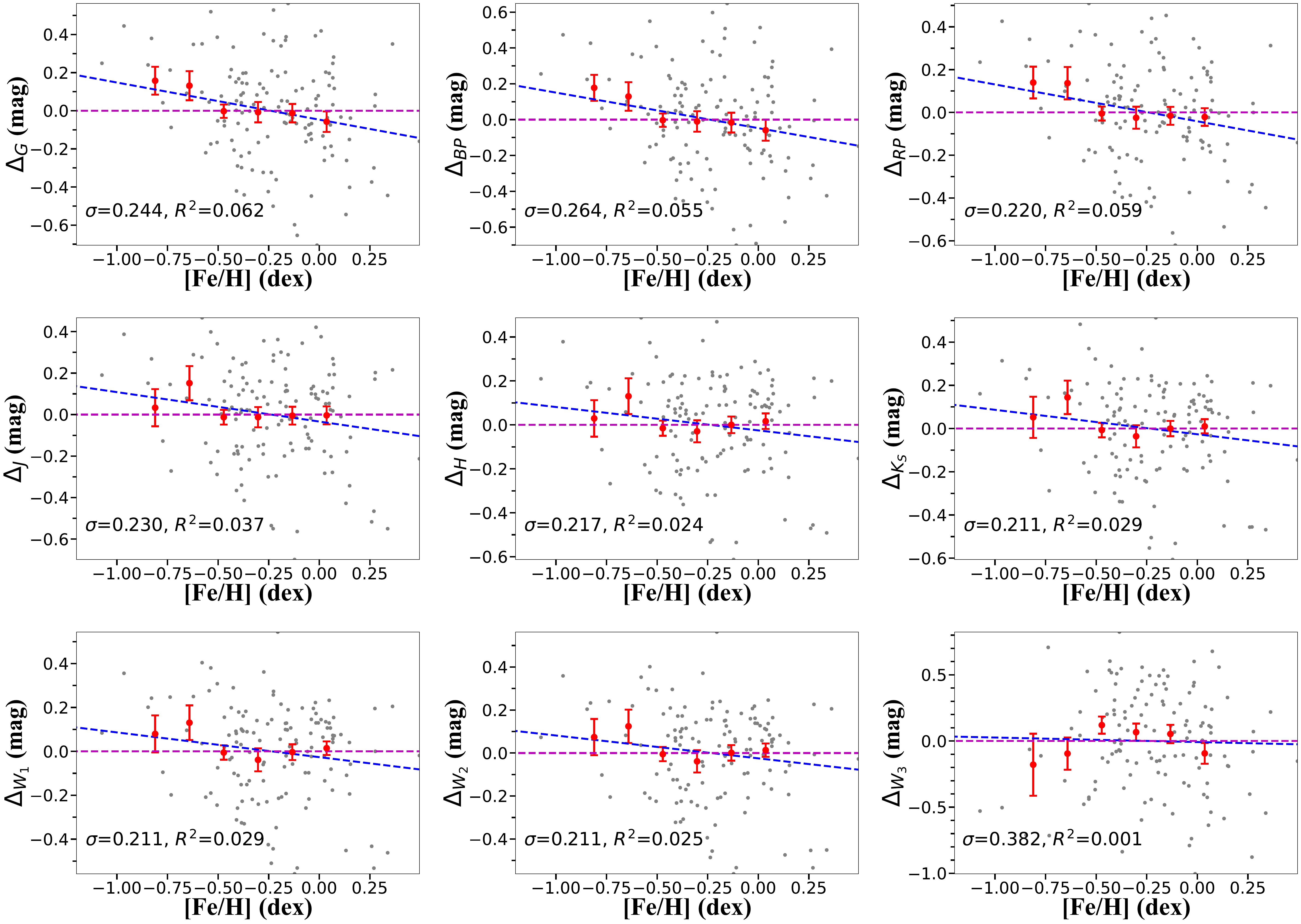} 
\caption{Residual plot for the PLR. The black dots show the distribution of the PLR residuals (the difference between the observed absolute magnitude and the fitted absolute magnitude) for each photometric band with respect to the metallicity. The red points and error bars represent the mean and standard deviation in different metallicity intervals. The blue dashed line is the fitted line for the black dots and the red dashed line is the 0-value line.} 
\label{f6} 
\end{figure}

\begin{figure}[ht!] 
\centering \includegraphics[width=1\textwidth]{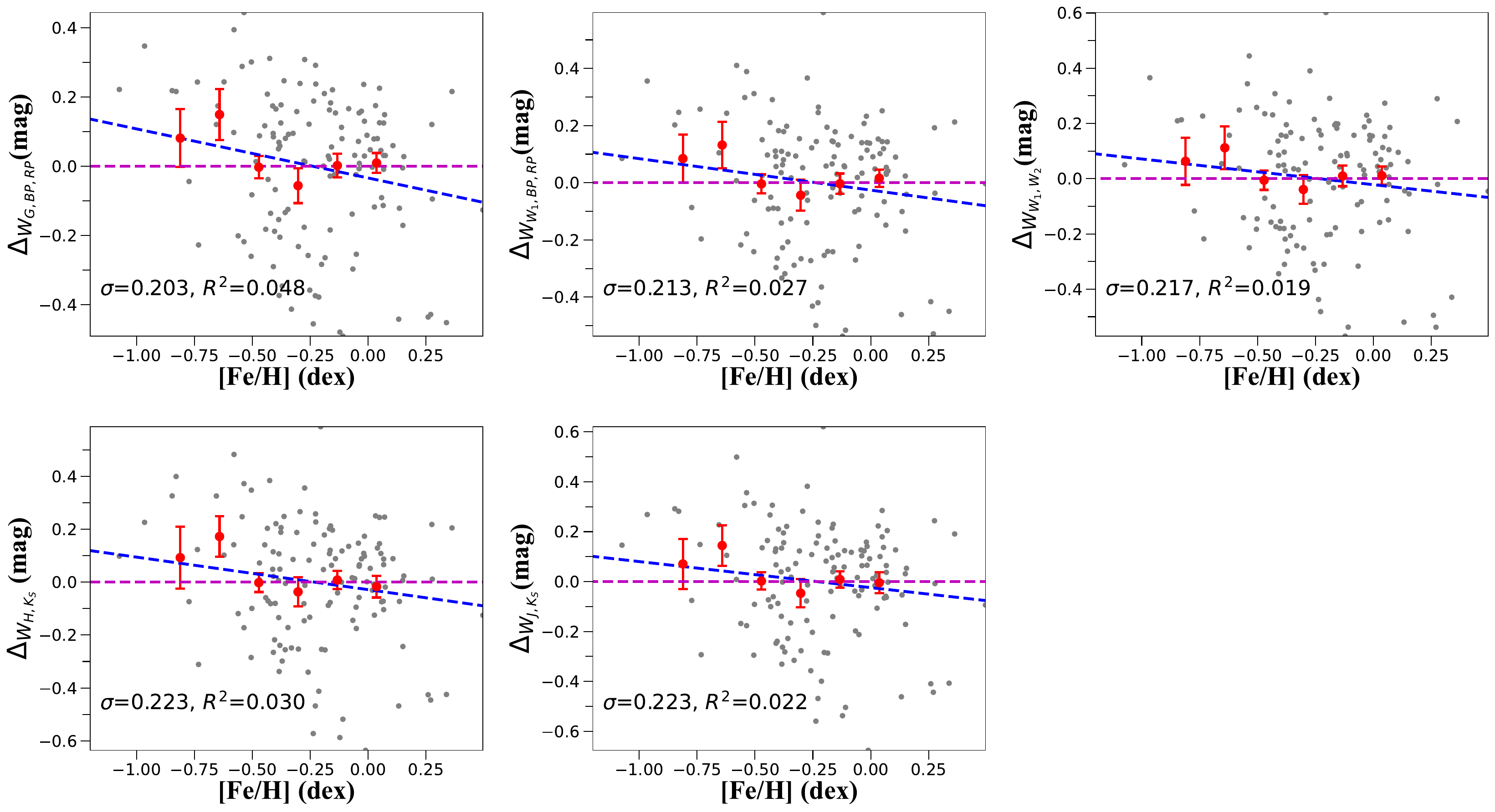} 
\caption{Similar to Figure \ref{f5}, but the residual plot for the PW relation.} 
\label{f7} 
\end{figure}

In Figures~\ref{f6} and~\ref{f7}, the red dots represent the correspondence between the average metallicity and the average absolute magnitude residuals for each interval. During the analysis, we found that the number of samples in some intervals was fewer than 5, which could lead to a significant increase in errors. Therefore, the red dots for these intervals are not displayed in the figures. The error bars in the figures show the $1\sigma$ dispersion within each metallicity interval, illustrating the range of residual values across different metallicity groups. By analyzing the residual plots, we found that the $R^2$ values for all bands are less than 0.07, indicating that the influence of metallicity on the PLR residuals is weak. We observed that if two points with an average metallicity of $-0.6$\,dex are excluded, metallicity has almost no significant impact on the PLR residuals. For $\delta$ Sct stars with metallicities less than $-0.6$\,dex, the deviation in the PLR residuals may be due to the small sample size, which is also supported by the larger dispersion. The analysis of the influence of metallicity on the PLR residuals once again demonstrates that the PLR of $\delta$\,Sct stars is weakly affected by metallicity, which is consistent with the large errors in the metallicity coefficients of the PLZ relation discussed in Section~\ref{sec:4}.

\begin{figure}[ht!] 
\centering \includegraphics[width=1\textwidth]{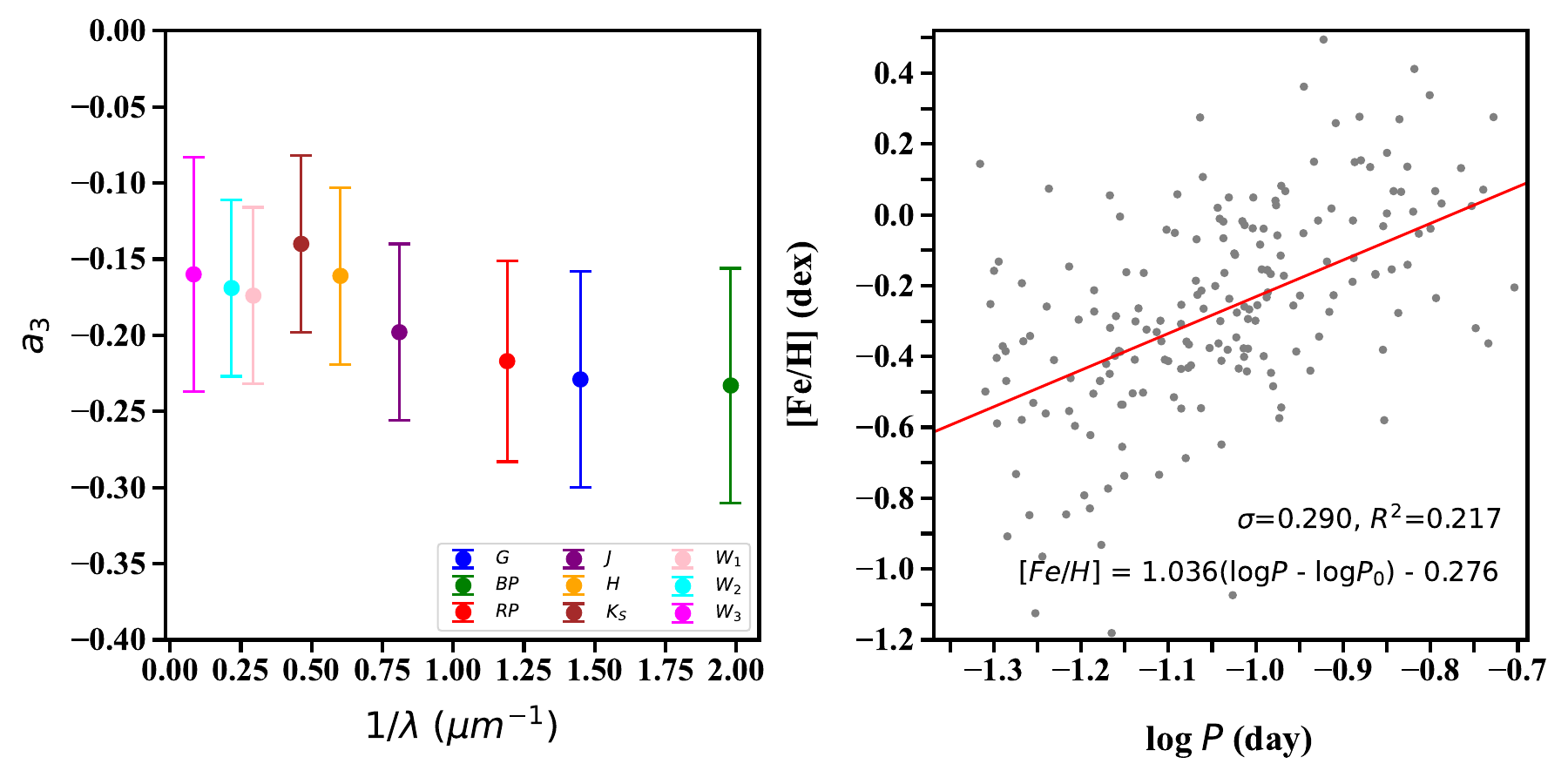} 
\caption{Relationship between the metallicity coefficient $a_3$ derived in Section \ref{sec:4} and the reciprocal of the wavelength ($1/\lambda$) (left panel). The correlation between metallicity and $\log P$ is shown in the right panel.}
\label{f8}
\end{figure}

In Section \ref{sec:4}, we derived the PLZ and PWZ relation for $\delta$ Sct stars and obtained the coefficients associated with the metallicity term. The left panel of Figure~\ref{f8} illustrates the relationship between these coefficients and the reciprocal of the corresponding wavelengths. As shown in the figure, the absolute values of the metallicity coefficients change minimally with decreasing wavelength. The influence of metallicity on the PLR in $\delta$ Sct stars exhibits characteristics distinctly different from those observed in Cepheids and RR Lyrae stars \citep{2024A&A...681A..65T,2015ApJ...808...50M,2023NatAs...7.1081C}. Specifically, the PLR of $\delta$ Sct variables is not significantly affected by metallicity, and the differences across various bands are not pronounced. In contrast, the metallicity significantly impacts the PLR of Cepheids and RR Lyrae stars, with considerable variations observed across different bands. We speculate that this phenomenon primarily arises because $\delta$ Sct variables possess higher temperatures, resulting in fewer metallic lines compared to Cepheids and RR Lyrae stars.

\cite{2020MNRAS.493.4186J} observed a correlation between the pulsation period and the metallicity of $\delta$ Sct stars. In our study, we conducted a similar analysis on our dataset. Equation \ref{equ9} presents the fitting equation we derived, and Figure \ref{f8} (right panel) illustrates the relationship between metallicity and $\log P$ based on our sample. Notably, the coefficient of determination obtained from our fitting is very low ($R^2 = 0.217$), indicating that the influence of metallicity on the period of $\delta$ Sct stars is not significant. Combining this with our previous analysis, we conclude that while metallicity does have some effect on the PLR, this effect is not substantial.
\begin{equation}\label{equ9}
    \text{[Fe/H]} = (1.036\pm0.142) (\log P - \log P_0) - (0.217\pm0.021),
\end{equation}

\subsection{Comparison of the PLR with the literature}
\begin{figure}[ht!] 
\centering \includegraphics[width=1\textwidth]{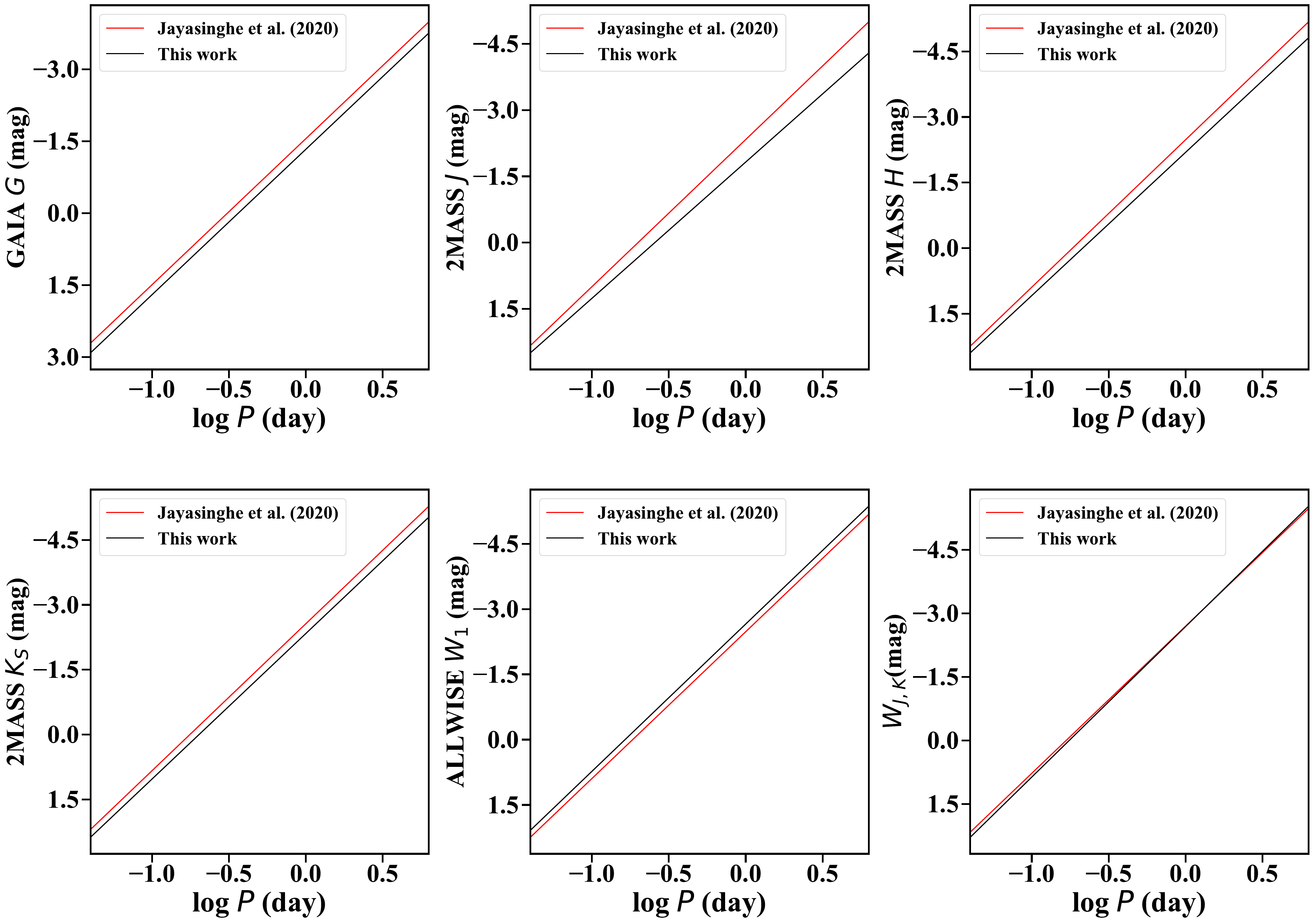} 
\caption{Comparison of the PLR for $\delta$ Sct stars in six bands from this work with the results from \cite{2020MNRAS.493.4186J}. The black solid line represents the fit derived in this study.}
\label{f9}
\end{figure}

We compared the PLR derived in this study with the PLR proposed by \cite{2020MNRAS.493.4186J}. As shown in Figure~\ref{f9}, we plotted the PLR for the fundamental mode $\delta$ Sct stars in six bands from \cite{2020MNRAS.493.4186J} and compared them with our own findings. Table~\ref{tab2} shows a comparison of the coefficients for PLRs. The consistency between our results and those of \cite{2020MNRAS.493.4186J} suggests that the PLR for $\delta$ Sct stars remains robust across different datasets and methodologies. Compared to their work, we used Gaia DR3 parallax and carefully corrected it, so our PLR will be a bit more accurate. Considering the weak effect of metallicity on the PLR and the consistency of the PLR obtained from different samples, we prefer to use the PLR to calculate the distance of $\delta$ Sct stars rather than the PLZ relation.
\begin{table}
\centering
\caption{Comparison of Six-Band PLR parameters with \cite{2020MNRAS.493.4186J}.}
\label{tab2}
\begin{tabular}{ccccccc}
\hline
Band & \multicolumn{2}{c}{\(\sigma\)} & \multicolumn{2}{c}{\(a_1\)} & \multicolumn{2}{c}{\(a_2\)} \\ 
\cline{2-3} \cline{4-5}\cline{6-7}
 & J20 & This work & J20 & This work & J20 & This work\\ 
\hline
\(G\)    & $0.210$ & $0.252$ & $-3.047 \pm 0.555$ & $-3.003 \pm 0.155$ & $-1.554 \pm 0.301$ & $-1.332 \pm 0.161$\\ 
\(J\)    & $0.188$ & $0.235$ & $-3.330 \pm 0.064$ & $-3.084 \pm 0.145$ & $-2.333 \pm 0.079$ & $-1.824 \pm 0.151$\\ 
\(H\)    & $0.185$ & $0.219$ & $-3.373 \pm 0.037$ & $-3.277 \pm 0.135$ & $-2.479 \pm 0.060$ & $-2.191 \pm 0.141$\\ 
\(K_S\)  & $0.183$ & $0.214$ & $-3.397 \pm 0.044$ & $-3.364 \pm 0.132$ & $-2.563 \pm 0.064$ & $-2.337 \pm 0.137$\\ 
\(W_1\)  & $0.190$ & $0.214$ & $-3.385 \pm 0.036$ & $-3.386 \pm 0.132$ & $-2.479 \pm 0.036$ & $-2.656 \pm 0.137$\\ 
\(J, K_S\) & $0.184$ & $0.216$ & $-3.475 \pm 0.034$ & $-3.555 \pm 0.139$ & $-2.699 \pm 0.004$ & $-2.690 \pm 0.145$\\ 
\hline
\end{tabular}
\end{table}
\section{CONCLUSIONS}\label{sec:6}
In this study, we integrated the $\delta$ Sct star catalogs from ZTF and ASAS-SN, metallicity data from LAMOST, and parallax measurements from Gaia to optimize the PL and PLZ relation. To mitigate the potential impact of the Gaia parallax zero point offset on the PLR, we performed a correction. Using the three optimal bands, $W_1$, $W_2$, and $W_{G,BP,RP}$, we determined the optimal zero point value as $zp_\varpi = 35 \pm 2 \, \mu\text{as}$. To derive the optimal PLR, we varied the parallax error threshold while fixing the Gaia parallax zero point and found that the total error in the PLR zero point was minimized at a parallax threshold of 6\%, with a corresponding error of 0.9\%. Subsequently, we established PL and PLZ relation across nine different bands, ranging from optical to mid-infrared, as well as five Wesenheit bands. Our results are consistent with previous studies, confirming the robustness and reliability of the PLR across different datasets and methodologies. 

By introducing a metallicity term into the PLR, we analyzed the relationship between metallicity intervals and PLR residuals, the correlation between metallicity term coefficients in each band, and the relationship between metallicity and $\log P$. We found that metallicity has a negligible impact on the PLR for $\delta$ Sct stars, with minimal variation across different bands. The metallicity effect in the PLR of $\delta$ Sct stars is weaker compared to that of Cepheids and RR Lyrae stars. Based on the current data, we conclude that the distances of $\delta$ Sct stars can be adequately determined using the PLR alone.

As the quality of data continues to improve, particularly with ongoing and upcoming astrometric and spectroscopic survey projects, our understanding of the PLR of $\delta$ Sct stars will be further enhanced. The Wide Field Survey Telescope, with its large aperture and wide field of view, is capable of observing large samples of short-period $\delta$ Sct stars across the Galactic disk \citep{2023SCPMA..6609512W, 2024arXiv241212601L}. Consequently, these stars will be more effectively utilized as tracers for distance measurement and structure studies.

\normalem
\begin{acknowledgements}
We thank the anonymous referees for the helpful comments. This research was supported by the National Natural Science Foundation of China (NSFC) through grants 12173047, 12373035, 12322306, 12373028, 12233009, 12133002. X. Chen and S. Wang acknowledge support from the Youth Innovation Promotion Association of the Chinese Academy of Sciences (CAS, No. 2022055 and 2023065). We also thanked the support from the National Key Research and development Program of China, grants 2022YFF0503404. We used data from European Space Agency mission Gaia (\url{https://www.cosmos.esa.int/gaia}), processed by the Gaia Data Processing and Analysis Consortium (DPAC, \url{ https://www. cosmos.esa.int/web/gaia/dpac/consortium}). Funding for the DPAC has been provided by national institutions, in particular the institutions participating in the Gaia Multilateral Agreement. This work has made use of LAMOST data. Guoshoujing Telescope (LAMOST; \url{http://www.lamost.org/public/}) is a National Major Scientific Project built by the Chinese CAS. Funding for the project has been provided by the National Development and Reform Commission. LAMOST is operated and managed by the National Astronomical Observatories, CAS.
\end{acknowledgements}
  
\bibliographystyle{raa}
\bibliography{ref}

\end{document}